\pdfoutput=1
\documentclass[fleqn,usenatbib]{mnras}

\usepackage{listings}
\usepackage[T1]{fontenc}

\DeclareRobustCommand{\VAN}[3]{#2}
\let\VANthebibliography\thebibliography
\def\thebibliography{\DeclareRobustCommand{\VAN}[3]{##3}\VANthebibliography}

\usepackage{graphicx}	
\usepackage{amsmath}	
\usepackage{amssymb}	
\usepackage{xcolor}     
\usepackage{paralist}   
\usepackage{fancyvrb}   
\usepackage{newtxtext,newtxmath}
\usepackage{tikz}
\usetikzlibrary{arrows}
\tikzstyle{block} = [draw,fill=blue!20,minimum size=2em]
\newcommand{\Hermes}{\textsc{Hermes}}

\title[Introducing Hermes: Fast CQL Execution]{Introducing Hermes: Executing Clinical Quality Language (CQL) at over 66 Million Resources per Second (inexpensively)}

\author[Kastroulis, Bonfini, Litsas]{
Angelo Kastroulis,$^{\ddag}$
Paolo Bonfini,$^{\ddag}$
Anastasios Litsas $^{\ddag}$
\\
$^{\ddag}$Ballista Technology Group \\ \{angelo.kastroulis, paolo.bonfini, anastasios.litsas\}@ballista.com
}

\pubyear{2022}

\begin{document}
\label{firstpage}
\pagerange{\pageref{firstpage}--\pageref{lastpage}}
\maketitle

\begin{abstract}
Clinical Quality Language (CQL) has emerged as a standard for rule representation in Clinical Decision Support (CDS) and Electronic Clinical Quality Measurement (eCQM) in healthcare. 
While open-source reference implementations and a few commercial engines exist, there is still a market need for high-performance engines that can execute CQL queries on the scales of millions of patients.
We introduce the \Hermes{} engine as the world's fastest commercial CQL execution engine.
\end{abstract}




\begingroup
\let\clearpage\relax
\tableofcontents
\endgroup

\section{Introduction}

Clinical Decision Support (CDS) provides the right information to the right decision maker \textit{at the right point of decision time} \cite{healthit.gov_2018}. Electronic Clinical Quality Measurement (eCQM), on the other hand, is a mechanism for measuring and assessing the outcomes and process \textit{after the fact} \cite{healthit.gov_2013}. CDS informs \textit{before} something happens, and eCQM measures \textit{what happened}. Thus, they are two sides of the same process.

Computationally, CDS and eCQM are somewhat fundamentally at odds. CDS requires low-latency for intervention at the point of care, while eCQM requires execution at very high Throughput for measurement of clinical results. eCQM rule authoring relies on \textit{a priori} knowledge in their creation. One may be tempted to think that the time it takes to execute against a population is irrelevant in a measurement scenario, however the reality is that computations on conventional systems can take days. It would be far more advantageous to execute in seconds, allowing organizations to inspect and adapt for better outcomes. In the time between reporting periods, a variety of technical, organizational, and care changes may have come into play. The results of the reports become stale and unactionable almost as soon as they are computed. 

An ideal approach would allow a single solution to compute eCQM \textit{and} replace aging rules systems (such as prior authorization). Custom rules could be written in CQL or another language (such as SQL) and run on the same platform.

Big data has been characterized in terms of (at least) 4 `V's --- \textit{volume}, \textit{velocity}, \textit{variety}, and \textit{veracity}. This is especially applicable to health data \cite{watson}.  Big data solutions require a high level of computer science expertise to operate effectively. The healthcare informatics requires a high level of expertise in the domain and relevant standards. That presents a challenge to Health IT experts who must possess both skills to build expensive systems. Health IT-specific methods of encoding rules such as CQL have also emerged, but they are not inherently able to operate at the scale of big data solutions. Thus, both of these approaches suffer from a disparity between the knowledge required to operate each of them.

CQL has proven effective in quality reporting initiatives in the United States, as well as in various clinical settings  \cite{nguyen2019implementation,brandt2022design,braunstein2022public}. But, CQL only works with structured data, while around 80\% of data is unstructured \cite{ali2019survey,tekli2016overview}. An ideal solution would leverage the capability of big data tools to allow options for unstructured data and machine learning.

The \Hermes{} engine bridges the gap between the domains by compiling CQL into code that is executable on big data and streaming pipelines, while optimizing and encoding health-specific knowledge directly into the computations, allowing superior performance. Moreover, by allowing for a variety of execution runners, \Hermes{} also decouples from the underlying technology.



\begin{figure*}
    \includegraphics[width=0.6\textwidth]{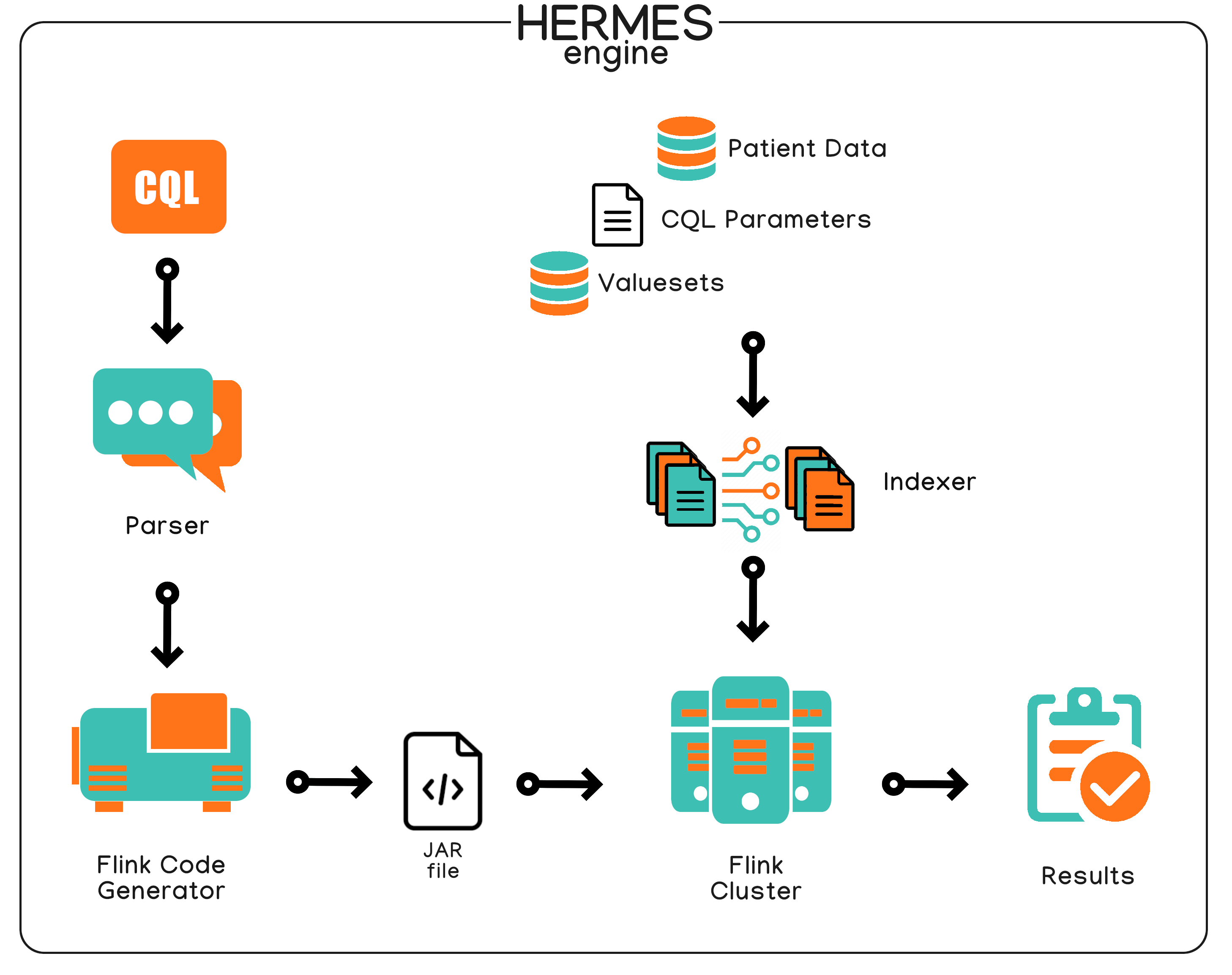} 
     \caption{
        Schematic representation of the rationale of the \Hermes{} engine.
        \label{fig:hermes}
     }
\end{figure*}

\section{Background and Related Work}
\label{sec:relwork}

Rule-based approaches (such as RETE) require that rules and data fit into working memory. As rule sets and data
increase, the amount of memory required increases more than linearly (a phenomenon known as “memory explosion”) \cite{chen2016dress}. Eventually, the working set becomes sufficiently large and the system breaks down. Similarly, small, fast interpreters fail when data sets become exceedingly large \cite{zaharia2010spark,kreps2011kafka}. Such algorithms are not appropriate for execution against large populations, making them ineffective on eCQM and rules executed against many patients.

Technologies in the big data ecosystem such as Apache Spark, Kafka, Flink, and Cassandra execute computations on large sets of data by partitioning the data and moving the computation to the data \cite{ali2019survey,zaharia2010spark,kreps2011kafka}. Data movement has become the new bottleneck and open research shows that systems that carefully minimize movement perform better \cite{kastroulis2019towards}. Apache Flink has emerged as one of the fastest computational platforms available \cite{verbitskiy2016use,bergamaschi2017bigbench}. 

Decoupling the underlying execution technology from the query engine (as it is done by Apache Beam \cite{room2020apache}) allows for a variety of ``runners''. In addition to decoupling, the next generation of systems will interpret queries by their intent and optimize accordingly \cite{kersten2011researcher}. These systems should not only work with batch loads, but also streams. Big data streams are generated continuously at unprecedented speed, yet allocating the cloud resources necessary to make them successful has emerged as a major research problem \cite{kaur2017efficient}.

Ideally, one would compile CQL into some code that can run on top of another widely-accepted computational platform (runner). Such a solution would also meet the needs of optimizing stream \textit{vs} batch workloads. Additionally, purpose-built systems are faster than general purpose systems because they can exploit certain characteristics of their domain that general-purpose systems cannot. For example, Health data encodes semantic meaning within it and has standardized components which are immutable and hence represent good candidates for optimization (e.g., a versioned, standardized Value Set). Whereas a general-purpose system is not able to infer this from a generic structure.

While open source CQL execution engines do exist, they suffer from serious performance problems because they are not horizontally scalable. Horizontal scalability is essential in reducing costs and increasing performance since vertical systems quickly run into high costs and hardware limitations \cite{ali2019survey}. Those engines' only recourse is micro-optimization, but that cannot achieve improvements of orders of magnitude.

A purpose-built CQL compiler/execution engine that has the characteristics mentioned above would be far superior to other systems. In the following section we will demonstrate how \Hermes{} characterizes such a system by means of realistic workloads.

\section{Methods}
\label{sec:methods}

Our \Hermes{} engine is structured as depicted in Figure \ref{fig:hermes}: the CQL query is parsed in order to provide a proper input for the Flink Code Generator, which creates the code that is in turn used to spun the Flink Cluster.
This cluster does the actual heavy job by utilizing as an input the CQL Parameters,  the Valuesets, and the Patient Data.

\begin{table}
	\centering
	\caption{Table of Workloads. Columns represent \textit{millions of records} for each Workload id (e.g., Workload 100M has 100 million patients and 2.25 billion resources in the set).}
	\label{tab:workloads_table}
	\begin{tabular}{lcccc} 
		\hline
		 & 1M & 10M & 50M & 100M\\
		\hline
		Patients & 1 & 10 & 50 & 100 \\
		Conditions & 2 & 20 & 100 & 200 \\
		Encounters & 3 & 30 & 150 & 300 \\
		Medications & 5 & 50 & 250 & 500 \\
		Procedures & 1 & 10 & 50 & 100 \\
		Observations & 5 & 50 & 250 & 500 \\
		Coverages & 5.5 & 55 & 275 & 550 \\
		Total & 22.5 & 225 & 1,125 & 2,250 \\
		\hline
	\end{tabular}
\end{table}

To test \Hermes{}' computational ability, we decided to assess its performance when evaluating a quality measure chosen from the HEDIS 2022 FHIR set. The HEDIS measures are a set of standard rules that hospitals and payers must report annually. We selected BCS for its diverse workload (complex temporal logic) and variety of resources consumed, but was also a fair representation of the "average" complexity.
In order to assess the engine under a variety of situations, we considered different \emph{workloads}, as shown in Table~\ref{tab:workloads_table}.  Moreover, we also considered different \emph{configurations}, corresponding to different hardware capabilities (see Appendix~\ref{tab:configurations_table}).
Therefore, in practice, we run a grid of tests where the engine was executed for each combination of \emph{Workload} and \emph{configuration}.

\begin{minipage}{\columnwidth}
    \begin{lstlisting}[caption=Example of implementation for the BCS CQL,label={lst:cql-code},basicstyle=\tiny\ttfamily]
    define "Numerator":
      exists ([Observation: "Mammogram"] m
          where m.effectiveTime 
            ends during <time window>)
      
    define "Denominator":
       AgeInYearsAt(date from 
        end of <time window>
      )in Interval[52, 74]
        and Patient.gender.value = 'female'
        and <Member coverage in 
            time window with only allowable gaps>
        
    define "Exclusions":
      exists([Encounter: "Hospice Encounter"] e 
        where e.status='completed' 
            and <in time window>)
      or exists([Procedure: "Hospice Intervention"] p
        where p.status='finished' 
            and <in time window>)
      or exists([Procedure: "Mastectomy"] p 
        where p.status='completed')
      ...
      or exists([Condition: "Absence of Breast"] c 
        where c.prevalencePeriod 
            <in time window>)
    \end{lstlisting}
\end{minipage}

\Hermes{} has both a \textbf{streaming} and \textbf{batch} mode, but tests were performed in batch mode to determine computing power and Throughput achievable on 100 million patients and 2.25 billion resources (observations, conditions, medications are all resources). The engine must have sufficient work so that it has substantial computation (we want it to work intensively). That is measured in two dimensions: 1) \textit{difficulty} of the computation and 2) large enough \textit{selectivity}. Selectivity refers to the number of matches in a computation (in this case, the number of matching patients in the set). \Hermes{} excels at eliminating non-matching records using its internal indexing scheme and Flink's partitioning and worker semantics. A more naive approach would require each record to be unpacked, deserialized, and computed only to find it does not match. \Hermes{} recognizes immediately that a record does not have a matching code, for example. A set with small selectivity (100 million patients with 1\% matches) would return too quickly, playing into the \Hermes{}'s strength. While this would show a massive performance advantage over other systems (three or more orders of magnitude) --- and low selectivity is a typical scenario --- it wouldn't tax the system sufficiently.

What makes \Hermes{} exceptionally well suited for execution on large sets is that it rewrites computations into relational algebra, then performs set operations on them (complete with predicate push-down and other optimizations). In order to see the effect of such optimizations, it is important to have exaggeratedly high selectivity. We chose an arbitrary value of 20\% for our test scenarios.

\medskip

\textbf{Let's explore an example of the engine's behavior when applied to a specific CQL search.}
A common function in CQL is \textit{retrieve}. Retrieve is essentially a join between ValueSets and another resource. For example, line 2 of Listing~\ref{lst:cql-code} is a join between \textit{Observations} and the \textit{Mammogram Valueset} on their respective \lstinline{code} properties. The Mammogram ValueSet can contain hundreds of breast cancer codes. In relational algebra, this would be known as a \textit{left outer join}. In order to perform this operation, Flink would need to load the smaller relation (Valuesets) into memory and stream the larger (Observations). Still, all matches would need to be remembered, which, over tens or hundreds of millions of matches, consumes gigabytes of RAM. 

\Hermes{}' VS HyperCache feature eliminates this by creating a join operation that pre-compiles Valusets and distributes them to each Flink Taskmanager, substantially reducing the memory footprint by \textbf{a factor of} 5--10$\times$.

\subsection{Configurations}

We run our test adopting different system setups on Amazon Web Services cloud (AWS), in order to explore the dependency of the engine's performance on the specifics of the hardware. In this context, a ``\textit{configuration}'' represents the set of parameters which uniquely describes a given setup.
A variety of configurations were explored, possessing different computational cores, virtual memory, etc. --- see configurations details in Appendix~\ref{sec:configurations}.

We chose EC2 instances that were optimized for compute and memory. We then varied the amount of resources given to each worker and varied the number of workers.

A \textit{Taskmanager} is an independently operating worker node that can run either on the same machine as other worker nodes, or on independent machines in a cluster. Taskmanagers are coordinated by a \textit{jobamanger}, whose primary responsibility is to distribute and collect work. Taskmanagers can themselves divide work into smaller components called \textit{slots}. \textit{Parallelism} refers to the number of slots available to each Taskmanager. Thus a cluster may have hundreds or thousands of slots (e.g., 10 Taskmanagers with 10 slots each yield a total of 100 slots cluster-wide).

\subsection{Rules}
\label{sec:rules}



\begin{figure}
  \centering
    \begin{tikzpicture}[>=latex']
        \node at (0, -1) (input1) {$i_{Encounter}$};
        \node[block] at (2,-1) (block1) {$f_1$};
        
        \node at (4, -1) (input2) {$i_{Condition}$};
        \node[block] at (6,-1) (block2) {$f_2$};
        
        \node at (6, -2) (input3) {$i_{Patient}$};
        \node[block] at (4,-2) (block3) {$f_3$};
        
        \node at (2, -2) (input4) {$i_{Procedure}$};
        \node[block] at (0,-2) (block4) {$f_4$};
        
        \node at (0, -3) (input5) {$i_{Procedure}$};
        \node[block] at (2,-3) (block5) {$f_5$};
            
        \draw[->] (input1) -- (block1);
        \draw[->] (block1) -- (input2);
        
        \draw[->] (input2) -- (block2);
        \draw[->] (block2) -- (input3);
        
        \draw[->] (input3) -- (block3);
        \draw[->] (block3) -- (input4);
        
        \draw[->] (input4) -- (block4);
        \draw[->] (block4) -- (input5);
        
        \draw[->] (input5) -- (block5);
        \draw[->] (block5.east) -- +(0.5,0);
    \end{tikzpicture}
    \caption{
        CQL \lstinline{Exclusions} executed as written by an interpreter, i.e., the scans are executed in sequence for some function $f$ given an input $i$.
        \label{fig:cqlunoptimized}
    }
\end{figure}
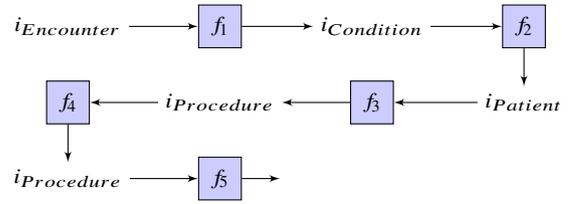

\tikzstyle{branch}=[fill,shape=circle,minimum size=3pt,inner sep=0pt]
\begin{figure}
  \centering
    \begin{tikzpicture}[>=latex']
    
        \node at (0, -1) (input1) {$i_{Encounter}$};
        \node[block] at (2,-1) (block1) {$f_1$};
        \draw[->] (input1) -- (block1);
        \draw[->] (block1.east) -- +(0.5,0);
        
        \node at (0, -2) (input2) {$i_{Condition}$};
        \node[block] at (2,-2) (block2) {$f_2$};
        \draw[->] (input2) -- (block2);
        \draw[->] (block2.east) -- +(0.5,0);
        
        \node at (0, -3) (input3) {$i_{Patient}$};
        \node[block] at (2,-3) (block3) {$f_3$};
        \draw[->] (input3) -- (block3);
        \draw[->] (block3.east) -- +(0.5,0);
        
        \node at (0, -4) (input4) {$i_{Procedure}$};
        \node[block] at (2,-4) (block4) {$f_4$};
        \draw[->] (input4) -- (block4);
        \draw[->] (block4.east) -- +(0.5,0);
        
        \node[block] at (2,-5) (block5) {$f_5$};
        \draw[->] (block5.east) -- +(0.5,0);
    
        \path (input4) -- coordinate (branch) (block4);
    
    
        \draw[->] (branch) node[branch] {}{ 
            } |- (block5);
    \end{tikzpicture}
    \caption{
        CQL \lstinline{Exclusions} optimized for parallelism for the same function $f$ given input $i$. Notice that the graph for Procedure has multiple ramifications because Procedure may be scanned multiple times.
        \label{fig:cqloptimized}
  }
\end{figure}
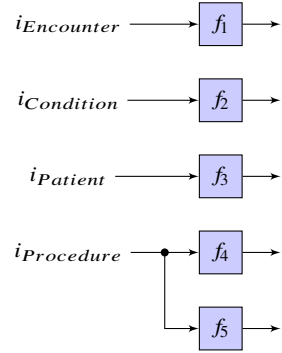



        
        

For this test, we selected the Breast Cancer Screening measure (BCS). A common convention for CQL quality measurement is to divide the top level rules into three main rules: \lstinline{Numerator}, \lstinline{Denominator}, and \lstinline{Exclusions}.  This specific nomenclature arises from the fact that quality measurements are intended to be reported as a ratio of the number of patients in which a target action was performed, over the size of the population eligible for that action. In other words, the proportion of patients that \textit{had} something done to those that \textit{should} have.

\textbf{For example, in the case of the breast cancer screening (BCS) introduced above} (see Listing~\ref{lst:cql-code}):
\begin{compactitem}
    \item \lstinline{Numerator} represents the number of women who had at least one mammogram within the previous 27 months (the time window),
    \item \lstinline{Denominator} is the number of women eligible (between the ages of 52 and 74 who visited the office),
    \item \lstinline{Exclusions} are the women who shouldn't be accounted for, presumably because they are already under treatment or  a care plan (e.g., women who had a mastectomy, are in hospice, have advanced illness, are in long term care, etc.).
\end{compactitem}

Instead of reporting the total ratio, however, the current state of the art is to pre-compute values atomically, allowing downstream systems to total them in a report.

In BCS, \lstinline{Numerator} is a set operation that scans observation codes within a time window. Since \Hermes{} indexes both time and codes, the operation computes extremely quickly.

\lstinline{Denominator} adds the complexity of a join between resources (\textit{coverage} and \textit{computing age}). The \textit{coverage} computation is particularly expensive since it aggregates coverage rows grouping, reducing, and eliminating entries in search of gaps. \lstinline{Exclusions} is the most complex and must scan a number of tables: \texttt{Procedures} (a few times), \texttt{Medications}, \texttt{Encounters}, \texttt{Conditions}, \texttt{Patients}, and \texttt{Observations}. Ultimately, \lstinline{Exclusions} performs a union and distinct operations on all of those tables for a positive match. Distinct Union is notoriously expensive because a system must remember the keys of the entire table. \Hermes{}' VS HyperCache was designed overcome this scenario. 

Executing the three operations together, while intensive, may also present an opportunity to reuse computation. For example, if a system could recognize that Procedures is scanned many times, it could combine the operations for substantial performance savings.

Most interpreter-based execution engines will translate the CQL to Expression Logical Model (ELM) and execute it sequentially as shown in Figure~\ref{fig:cqlunoptimized}.  ELM is a logical specification standard (an Abstract Syntax Tree), but it is not compact enough to execute directly in a performant manner. A better approach is shown in Figure~\ref{fig:cqloptimized}: the functions related to the resource scans could all be done at the same time on multiple slots. Note the $f_5$ and $f_6$ could actually operate on Procedures at the same time (or perhaps even become the same computation).

\subsection{Workload: Synthetic Data Generation}

So far we have presented the CQL rules and explained their rationale, but how were our tests performed, \textit{in practice}?

\begin{figure}
 	\includegraphics[width=\columnwidth]{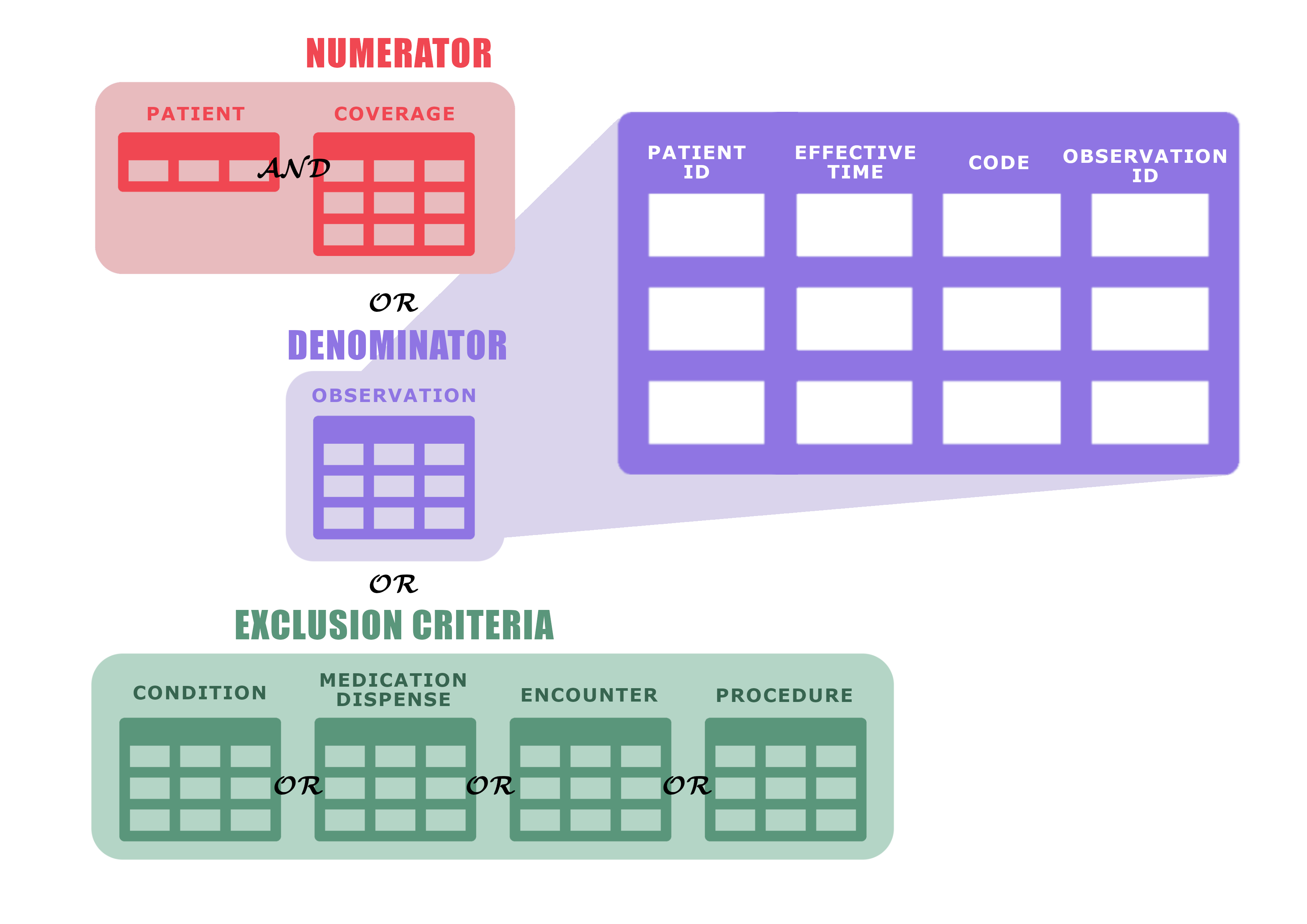}
     \caption{
         Structure of the synthetic dataset used for the performance test.
         The figure shows the entries for a single patient, which pertain to different tables, i.e., only the Patient table contains one single row per patient, all the other tables may contain multiple rows indexed by the same patient ID.
         The magnified table shows an example of the possible column contents. 
         The different tables are organized in 3 major groups: Numerator, Denominator, and Exclusion Criteria.
         The `\textsc{And}' and `\textsc{Or}' logical relations signify which conditions shall be satisfied for a match to happen.
         \label{fig:dataset}
     }
\end{figure}

First, we needed a dataset on which to apply the rules.
For this purpose, we needed be to run \Hermes{} against a synthetic dataset modeled over the structure of the fields expressed in the BCS measure (see Figure \ref{fig:dataset}).
This artificial dataset shall satisfy the following 2 properties. 
First, it contains between millions to hundreds of millions of synthetic patients; this is desirable because a large dataset provides a stress test that is sufficiently challenging for our engine.
Second, it should allow us to control the volume of matches produced by executing a given search (namely, the one in Listing~\ref{lst:cql-code}).  This is an extremely significant parameter, because an engine's workload scales, at the first order, with the number of expected matches.
One key factor in a successful engine design is the ability to quickly discard irrelevant information, focusing on the potential matches in order to save computational time.
What we sought was a database containing plenty of irrelevant information which would overload a sub-optimal engine.  As previously mentioned, we aimed at having about 20\% matches.

The generation of artificial structured (tabular) data is an active field of research \citep[e.g.][]{borisov2021deep}, which includes advanced techniques such as generative-adversarial networks \citep[GANs; e.g.,][]{abedi2022gan}.
Yet, to our knowledge, there exists no simple tool to generate data which can guarantee a preset number of matches given a [CQL] rule --- the most important structural factor we had to account for.
In fact, for the purposes of this test, it is not necessary for the data to be sampled according to a \textit{realistic} distribution. It is more important to determine whether results fell within/outside the matching ranges set by the search rules.

\medskip

Therefore, we ultimately resorted to creating our own data with a more direct approach.
Given that we wanted a 20\% match rate, we simply sampled random dataset entries such that they fell within/outside the search ranges accordingly.
We can categorize the synthetic entries into 2 types, according to this definition:

\vspace*{3pt}
\begin{tabular}{rl}
    \renewcommand{\arraystretch}{3}
    \textbf{valid} entry    & $\rightarrow$ within the \textit{allowed} range \\
    \textbf{in-valid} entry & $\rightarrow$ outside the \textit{allowed} range \\
\end{tabular}
\vspace*{3pt}

\noindent
where the allowed range is set by the rules (specified in $\S$\ref{sec:rules}).
Notice, though, that a valid entry does not automatically yield `a match'!
Because of the  database format of Figure \ref{fig:dataset}, a \text{match} is returned whenever there is a match between the search query and \textit{either} of the Denominator, Numerator, or Exclusion entries \textit{as a whole}.
More specifically:

\vspace*{3pt}
\begin{compactitem}
    \setlength\itemsep{2em}
    \item For the Denominator, 1 valid match is yielded when both the Patient and Coverage Tables are valid in all their entries

    \item For the Numerator and Exclusion, 1 valid match is yielded when either of their composing Tables is valid in all their entries
\end{compactitem}
\vspace*{3pt}

\noindent
The generation problem then becomes:

\vspace*{5pt}
\begin{minipage}{.4\textwidth}
    \textit{How shall we distribute valid and in-valid entries across the dataset, so that they collectively yield to matches for 20\% of the data volume?}
\end{minipage}
\vspace*{5pt}

\noindent
The number of data per patient are given by the sum of Denominator ($D$),
Numerator ($N$), and Exclusion ($E$) data per patient. 
Note, though, that we want to generate a variable number of such data.  Therefore, let's consider their \textit{average} numbers: $\mu_C$, $\mu_N$, and $\mu_E$, and estimate the \textit{data volume per patient}, $T_p$, as:

\noindent
\begin{align}
    T_p = \mu_D + \mu_N + \mu_E
\end{align}

\noindent
and, from there, derive the amount of \textit{valid data per patient} $V_{T_p}$
(proportional to the number of computations that will result in a match for
that patient), given a ratio $r$ of desired matches (e.g. 20\%):

\noindent
\begin{align}
    V_{T_p} & = r~T_p = r~(\mu_D + \mu_N + \mu_E)\nonumber\\
            & = r~\mu_D + r~\mu_N + r~\mu_E
    \label{eq:V_T_P}
\end{align}

Let's now recollect that Denominator, Numerator, and Exclusion are actually composed by multiple tables, each having a set of entries (their columns). E.g., Numerator is composed of a single \textit{Patient} ($P$) table, and a $Coverage$ table, both of which shall host valid entries to return a match.
We can therefore re-write Equation \ref{eq:V_T_P} as:

\noindent
\begin{align}
V_{T_p} & = r~\mu_D + r~\mu_N + r~\mu_E\nonumber\\
        & = r~(P\times\mu_C) + r~\mu_N + r~\mu_E\nonumber\\
        & = r~(\mathbb{1}\times\mu_C) + r~\mu_N + r~\mu_E
\end{align}

\noindent
where $\mathbb{1}$ is just a placeholder to remind us that even if we have 1 single \textit{Patient} table per patient, that must be actually sampled.
We can redistribute $r$ between \textit{Patient} and \textit{Coverage} tables:

\noindent
\begin{align}
V_{T_p} & = (\sqrt{r}~\mathbb{1}\times\sqrt{r}\mu_C) + r~\mu_N + r~\mu_E
\end{align}

\noindent
We can read this equation as:

\vspace*{5pt}
\begin{minipage}{.4\textwidth}``
    \textit{At generation time, we have to sample a matching Patient table row $\sqrt{r}$ of the times, a matching Coverage table row $\sqrt{r}$ of the times, and any of the $N$ or $E$ tables rows $r$ of the times''. } 
\end{minipage}
\vspace*{5pt}

We are just left with defining what is  a ``\textit{matching table row}'', but that is trivial: a row whose entries are all within the ranges allowed by the rules. In other words, when a row is to be generated:

\vspace*{3pt}
\begin{tabular}{rcl}
    \renewcommand{\arraystretch}{3}
    as matching              & $\rightarrow$ & all of its entries are sampled as valid \\
    as \textit{not} matching & $\rightarrow$ & at least 1 of its entries is sampled\\
                             &               & as in-valid. \\
\end{tabular}
\vspace*{3pt}

\medskip

With this rationale, we created 4 synthetic datasets of different sizes, hosting 1, 10, 50, or 100 million (respectively labelled $1M$, $10M$, $50M$, and $100M$) synthetic patients\footnote{
 Note that, for consistency, we made such that the larger datasets \textit{include} the smaller ones.
}.
In the remainder, we refer to these test datasets as `\textit{Workloads}'.




\subsection{Test Setup}

We chose Flink as the underlying runner and compiled BCS to a jar file, distributed on each cluster configuration. We stood up several cluster configurations (see Appendix~\ref{sec:configurations}) and executed the various Workloads (see Table~\ref{tab:workloads_table}) against each configuration.

We also tested the effect of a variety of index formats (Apache Avro, Orc, and Parquet). Finally, we compared executions with VS HyperCache enabled vs disabled.





\section{Observations and Test Results}

\begin{figure}
    \centering
     \includegraphics[width=0.75\columnwidth]{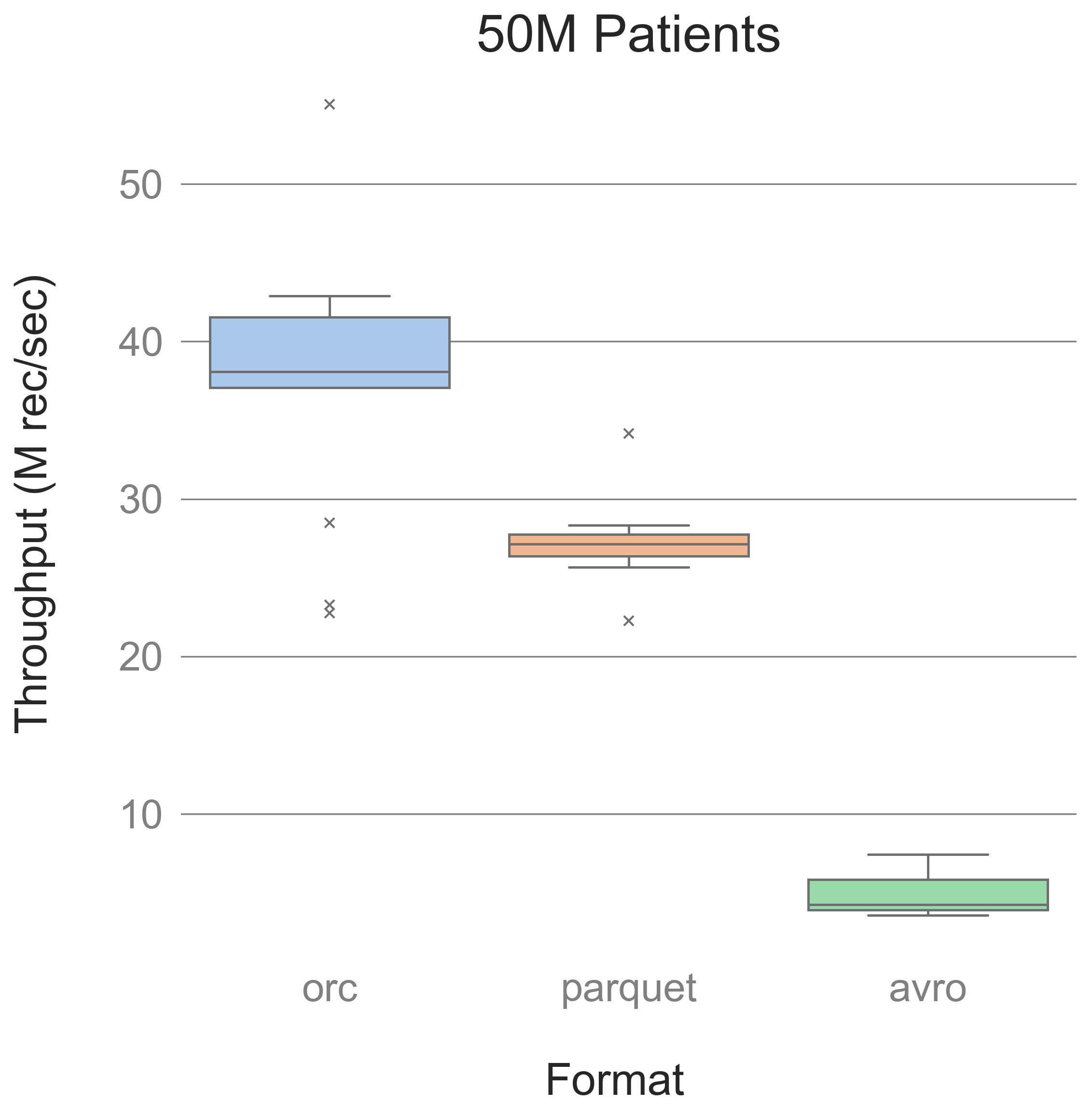}
     \caption{
         Comparing Throughput of the CQL computation using Apache Avro, Parquet, or Orc file types, the 50 million patient Workload illustrates the relative performance indicative of all Workloads we tested. In every Workload, Apache Orc was the fastest.
        \label{fig:formats}
     }
\end{figure}

\begin{figure*}
    \includegraphics[width=0.8\textwidth]{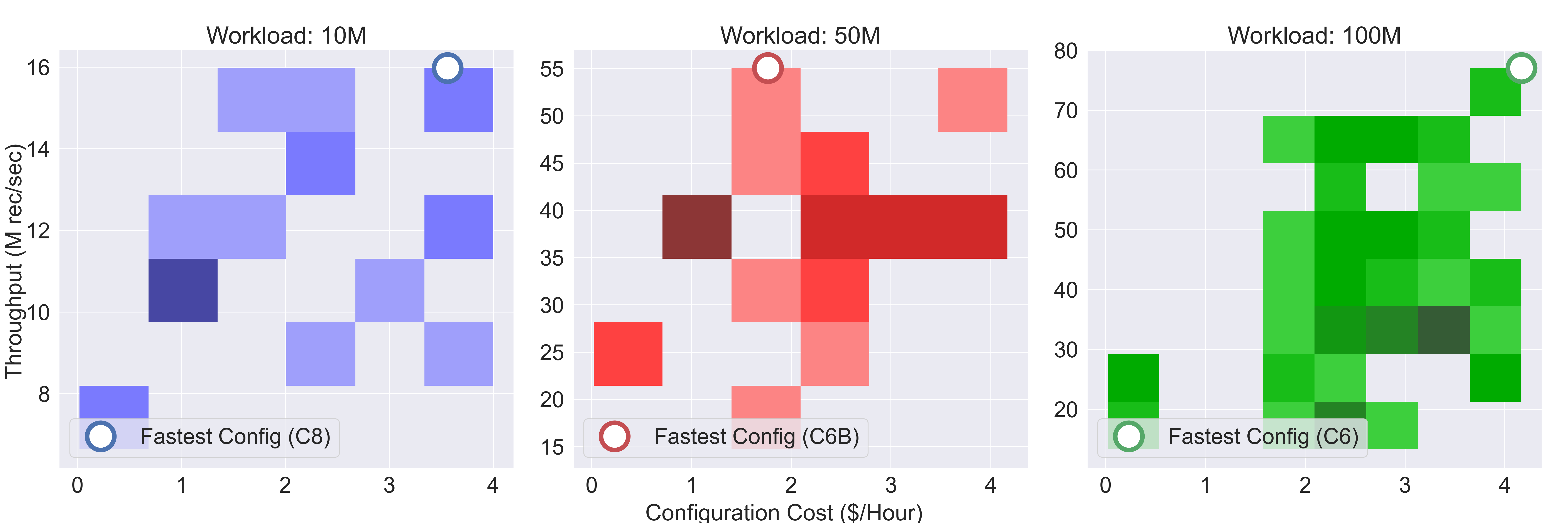}
     \caption{Comparing Throughput (millions of resources per second) and Cost (dollars per hour) of Workloads.
     From \textit{left} to \textit{right}, we show the results for increasing Workloads.
     The color gradient indicates the number of configurations that occupy that locus of the plot.  The data point indicates the best-performing configuration.
     \label{fig:throughput_vs_cost}
     }
\end{figure*}

\begin{figure*}
 	\includegraphics[width=0.8\textwidth]{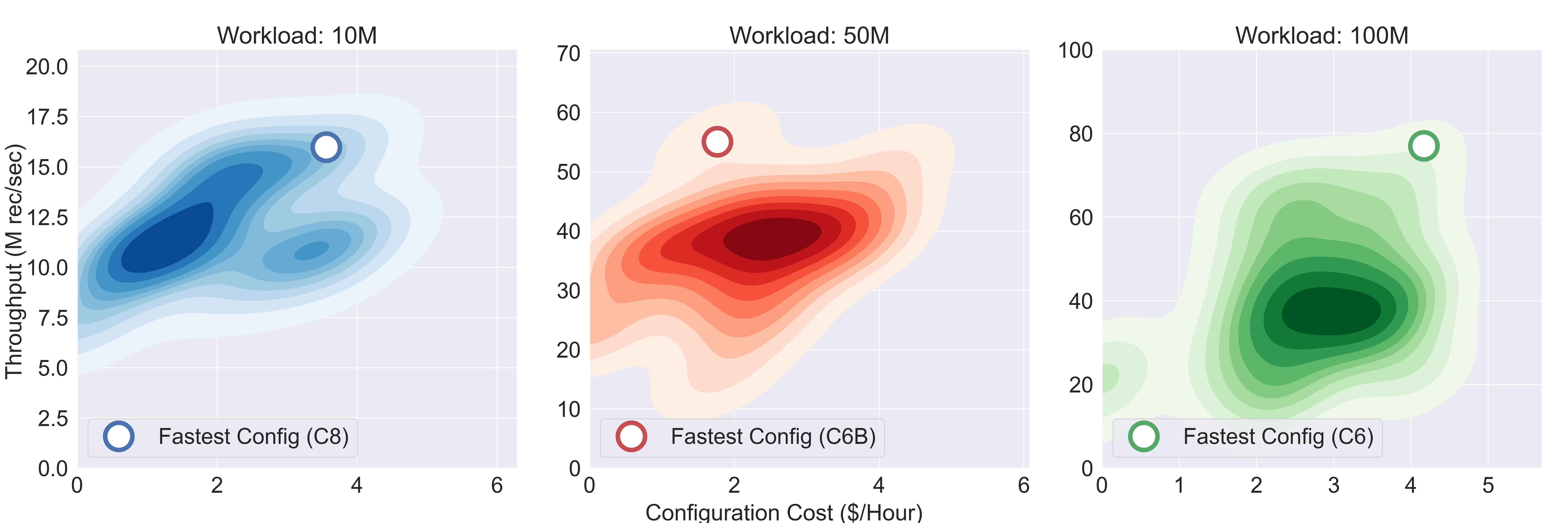}
     \caption{Comparing Throughput (millions of resources per second) and Cost (dollars per hour) of Workloads.
     From \textit{left} to \textit{right}, we show the results for increasing Workloads.
     The color gradient indicates the density of configurations that occupy that locus of the plot, obtained via a Kernel Density Estimator (KDE).  The data point indicates the best-performing configuration.
     \label{fig:throughput_vs_cost2}
     }
\end{figure*}

\begin{figure}
\centering
    \includegraphics[width=\columnwidth]{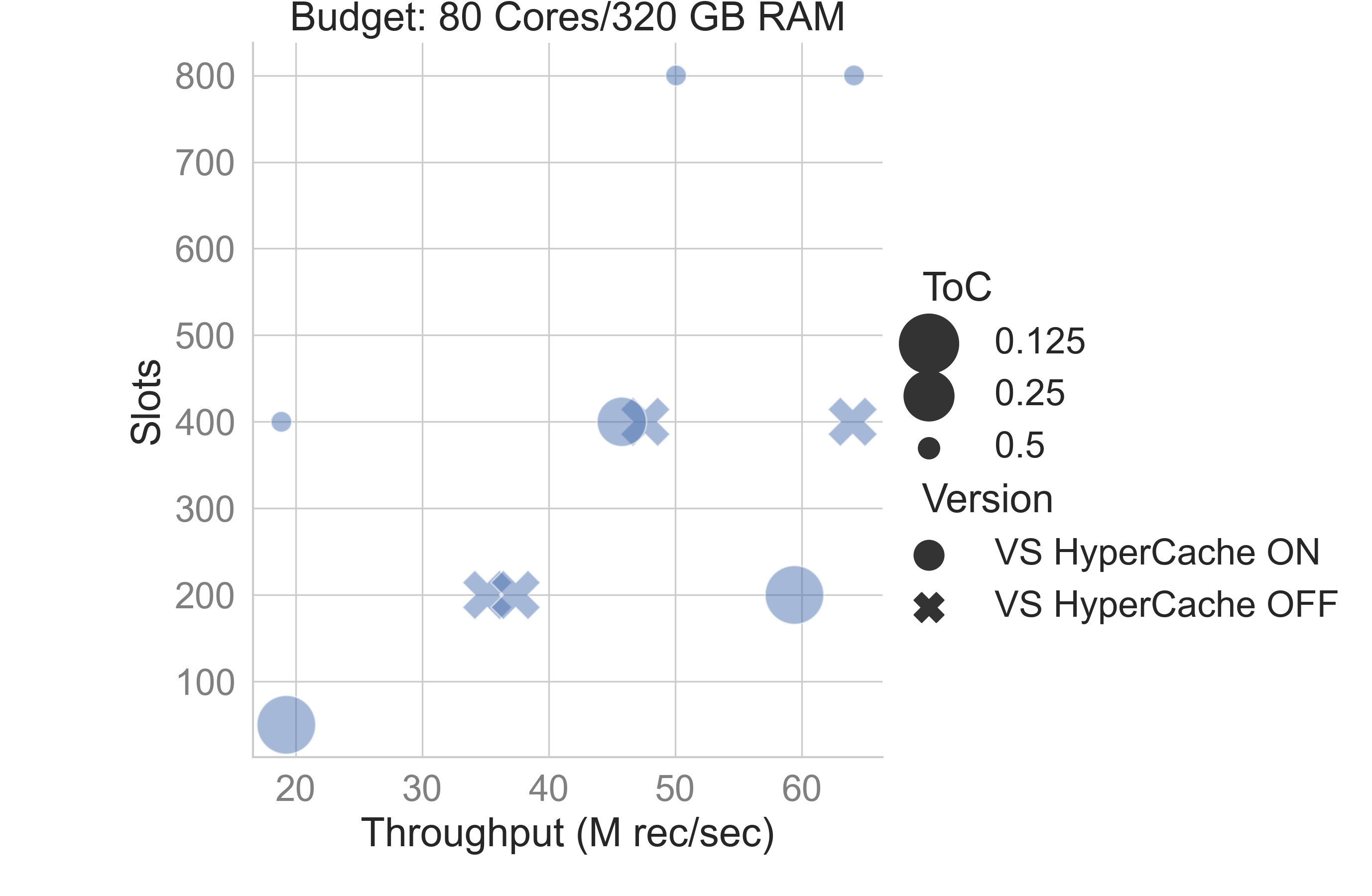}
     \caption{
         Comparing the distribution of resource budgets.
         The data points are size-coded based on the the ratio of Taskmanagers over Cores ($ToC$), which implies a larger (lower) distribution of work for higher (lowerBecause) values.
         The two different marker types refer to the activation (or not) of the HyperCache feature.
         \label{fig:resource_distribution}     
     }
\end{figure}

\subsection{Index Format}

\Hermes{} is capable of ingesting data in a variety of formats. During the compilation phase of CQL, it identifies which elements are referenced and indexes them (such as a patient's age, or a medication's code). These indexes are used at runtime rather than the original FHIR data for computations. This reduces the size of the data scattered across the cluster by two or more orders of magnitude. The index format can be any format specified through format connectors. We tested Apache Avro, Parquet, and Orc. Apache Avro has gained popularity through its use in Apache Kafka and has several advantages (such built-in schema definition, version evolution, and serialization/deserialization performance). Apache Parquet and Orc are columnar formats which allow advanced optimizations such as predicate push down. 

\bigskip

As can be seen in Figure \ref{fig:formats}, Apache Orc was the fastest in all cases, so from this point forward, we will exclude the other formats in discussion for simplification.

\subsection{Workloads}

\Hermes{}' advantage as a Throughput-optimized computational engine can be observed in Figure \ref{fig:formats}. As workloads increase (10M, 50M, and 100M), so does the Throughput, correspondingly (16, 55, and 66 million resources per second). The advantage of cost amortization of dividing and distributing work among a cluster pays off with more data. But, it also shows the advantage of applying predicates intelligently to reduce the amount of data moving between steps in the topology as early as possible. Engines commonly apply computations across the patient space, thus requiring traversal of all of the data because they never reduce the set. \Hermes{} rewrites the computation to work across the resources \textit{first} (medications, conditions, observations, etc.) reducing the work to be done at later stages.

\bigskip
\noindent
Thus, the more records to process, the faster \Hermes{} gets.

\subsection{Resource Arrangement: Budgets and Slots}

Within a specific \textit{budget} of resources (a given CPU, RAM, or other constraints), there are still decisions to be made as to how best to use them. For example, multiple configurations use a total cluster-wide consumption of 80 cores and 320 GB of RAM. How should such resources be distributed? Should we favor fewer, larger Taskmanagers (10 with 8 cores and 32GB of RAM each) or many, smaller ones (40 with 1 core and 8GB of RAM each)?

We label ``\textit{ToC}'' this relationship of Taskmanagers to resources (specifically, cores). A smaller ToC indicates relatively fewer, larger Taskmanagers, while a larger ToC indicates relatively more, smaller Taskmanagers.

\textit{Slots} are the logical execution cores that are parallelized in a computation. The total number of slots in a cluster are computed as $$n_{slots} = n_{parallelism} \times n_{taskmanagers}$$ Is it better to have a single slot per Taskmanager, or to have multiple slots per Taskmanager? 

Figure \ref{fig:resource_distribution} shows that the way we distribute the resources on the cluster (ToC) and the number of slots cluster-wide are related in the way they affect Throughput.  A smaller number of slots (200) performs well (about 56 million resources per second) on a smaller ToC (larger Taskmanagers), whereas increasing the number of slots (to 800) allows the larger quantity of smaller Taskmanagers to do more work. Note that at the extreme end of the chart lie the two fastest configurations, namely:
\begin{enumerate}
    \item 800 Slots, 0.5 ToC, VS HyperCache On
    \item 400 Slots, 0.25 ToC, VS HyperCache Off
\end{enumerate}

The first (with VS HyperCache enabled) allows for many more slots because the memory required for each is far less. The second could not run as many slots and Taskmanagers because each Taskmanager required at least 12GB of RAM. VS Hypercahce effectively reduced the amount of memory required per Taskmanager.

\bigskip

Many smaller Taskmanagers is preferable because the jobmanager may be able to schedule them for other rules sooner.  This happens because each of them is doing relatively less work, and hence they complete more quickly individually. VS HyperCases proves effective in reducing the memory footprint, allowing for higher parallelization.

\subsection{Cost}

\noindent
One key element of the performance analysis is estimating the \textit{cost} of the hardware setup used to execute an engine's search.  
This is trivial because given a task of any engine --- no matter how inefficient such an engine may be---, there will always be a hypothetically more expensive setup that can boost that engine's performance. 

\medskip

The cost of the configurations we tested was estimated by adopting the price-per-hour (\textit{CpH})  of the cloud service that was utilized  (Appendix~\ref{tab:characteristics_table}), and re-normalizing it by the resources that we actually allowed Flink to access (see Appendix~\ref{tab:configurations_table}).
For example, Image Type \textsc{m5.24xlarge} costs \$4.08 per hour with its 384~GB RAM and 96 cores, but e.g. configuration \textsc{C18B} --- based on the aforementioned \textsc{m5.24xlarge} --- uses only 16~GB RAM and 2 cores.

\medskip

We can think of the $CpH$ of a given Image Type $i$ as the weighted sum of its CPU and RAM resources times the unit cost $u_{i}$:

\noindent
\begin{align}
    CpH = ( \alpha N_{c, i} + N_{r, i} ) \times u_{i}
\end{align}

\noindent
where $N_{c, i}$ and $N_{r, i}$ are the CPU and RAM units in Image Type i, respectively, and $\alpha$ is the weighing factor that accounts for the different cost of a unit of 1~GB of RAM and 1 core.
This weighting factor can be calculated by comparing the market cost of a `building block' of RAM (64~GB) and one of CPU (8~cores); we roughly estimated $\alpha \sim 6$.

Notice that $u_i$ is the unit cost per CPU or RAM resource, regardless, for Image Type $i$.
So we can reverse the above formula, to get the unit cost $u_i$ (which is all we need to estimate the cost of a configuration):

\noindent
\begin{align}
    u_i =  {CpH  \over \alpha N_{c, i} + N_{r, i}}
\end{align}

\noindent
Following on our example for Image Type \textsc{m5.24xlarge}, Appendix~\ref{tab:characteristics_table} gives $CpH_i = \$4.608$/h,  $N_{c, i} = 96$, and $N_{r, i} = 384$, hence we obtain $u_{i} = \$4.608  / (6 \times 96 + 384 ) \sim \$0.005/h/resource$.

With $u_{i}$ at hand, we can calculate the cost-per-hour for any configuration ($CpH_{config}$) derived from Image Type $i$, which utilizes the actual resources $\bar{N_c}$ and $\bar{N_r}$\footnote{
    Do not confuse $\bar{N_c}$ and $\bar{N_r}$ with $N_{c, i}$ and $N_{r, i}$: the former refer to the configuration-restricted resources, the latter refer to the maximal capacity of Image Type $i$ that the configuration is built upon.
}.
simply as:
\noindent
\begin{align}
    CpH_{config} = ( \alpha N_c + N_r ) \times u_i
\end{align}

\noindent
For example, let's calculate $CpH_{config}$ for configuration \textsc{C18B}, which is based on Image Type \textsc{m5.24xlarge}.
From Appendix~\ref{tab:configurations_table} we have $N_c = 2$, $N_r = 16$, and using the previously calculated $u_i$, we 
obtain $CpH_{config} =$ $(6 \times 2  + 16 ) \times \$0.005 = \$0.14~/hour$.

\bigskip
\noindent
In summary, our exploratory testing mapped the performance distribution of diverse configurations  in the cost--Throughput space. Notably, we observe that the fastest configuration of Workload 50 M (indicated by a circle in Figures~\ref{fig:throughput_vs_cost} and \ref{fig:throughput_vs_cost2}) was \textit{also} one of the least expensive, at 1/3 the maximal experimented cost.  Reverting the argument, the relatively ``flat'' top of the distributions in Figure \ref{fig:throughput_vs_cost2} suggests that  --- with a smart choice of configuration --- we might reduce the cost by a factor of 3 and still retain $\sim$90\% of the performance of the best (and most expensive) configuration.

\section{Conclusions}

\subsection{Conclusion}

\textbf{Performance.} \Hermes{} performs exceptionally well with large data sets. In fact, the more data it is given, the more records per second it can compute because the cost of distributing the workload is quickly recovered. It is also extremely efficient at computation sharing.

\noindent
\textbf{Cost.} \Hermes{} provides an excellent trade-off between top performance and cost, operating at approximately 66 million resources computed per second at less than \$2 per hour. More aggressive pricing models (such as spot pricing) would cost far less (\$0.89 per hour). While not specifically tested, the cluster creation time in a kubernetes-style deployment (as is supported by \Hermes{}) allows new pods to be spun up nearly instantaneously, further reducing costs by resource-sharing.

\noindent
\textbf{Resource utilization.} Internal features such as predicate push down and VS HyperCache, proved extremely effective at reducing memory and compute consumption. \Hermes{} is highly parallelizable, working best with many small workers.

\noindent
\textbf{Eco-system friendly.} Finally, \Hermes{} is effective at using a variety of big data technologies such as Apache Flink, Orc, Kafka, Parquet, etc., and orchestration technology such as Kubernetes.

\subsection{Future Work}

The goal of this test was \textit{not} to tune \Hermes{}, but rather understand its performance characteristics and trade-offs. Undoubtedly, additional tuning (such as determining the optimal partitioning of data for maximizing parallelization) would have lead to higher Throughput or lower cost. What is the trade-off between partition size and Throughput?

Future tests should also measure the effect of executing \textbf{multiple rules simultaneously}, and the increased global Throughput resulting from combining rules on the same data in the same execution job. Additionally, durably caching \textbf{intermediate state} in future versions of \Hermes{} is also an area for improvement. Could performance be gained by analyzing the entire \textit{rule space} and determining which portions of data could be shared between them, thus amortizing the cost of caching to increase Throughput?




\bibliographystyle{ACM-reference-format}
\bibliography{bibliography} 



\newpage
\appendix
\onecolumn

\section{Table of Configurations}
\label{sec:configurations}
\label{tab:configurations_table}

	\centering
	\begin{tabular}{cccccccc} 
		\hline
		 ID & Image Family & Taskmanagers & Cores/TM & Ram/TM (GB) &  Parallelism & Network BW  & EBS  
                   BW \\
		\hline
		C1 & c6id & 2 & 1 & 12 & 10 & 12 & 10 \\
        C2 & c6id & 5 & 1 & 12 & 10 & 12 & 10 \\
        C3 & c6id & 10 & 1 & 8 & 10 & 37 & 30 \\
        C3b & c6id & 10 & 1 & 15 & 10 & 37 & 30 \\
        C4 & m5 & 10 & 1 & 32 & 10 & 25 & 19 \\
        C4B & m5 & 10 & 1 & 32 & 20 & 25 & 19 \\
        C4C & m5 & 10 & 1 & 32 & 5 & 25 & 19 \\
        C5 & c6id & 2 & 1 & 24 & 10 & 12 & 10 \\
        C6 & c6id & 5 & 1 & 30 & 10 & 37 & 30 \\
        C6B & c6id & 5 & 1 & 30 & 20 & 37 & 30 \\
        C8 & m5 & 20 & 1 & 16 & 10 & 25 & 19 \\
        C8B & m5 & 20 & 1 & 16 & 20 & 25 & 19\\
        C9 & c6id & 2 & 4 & 12 & 10 & 12 & 10 \\
        C10 & c6id & 5 & 4 & 12 & 10 & 12 & 10 \\
        C12 & m5 & 20 & 4 & 16 & 10 & 25 & 19 \\
        C12B & m5 & 20 & 4 & 16 & 20 & 25 & 19 \\
        C13 & c6id & 2 & 4 & 24 & 10 & 12 & 10 \\
        C13b & c6id & 2 & 12 & 24 & 10 & 12 & 10 \\
        C16 & m5 & 10 & 4 & 32 & 10 & 25 & 19 \\
        C16B & m5 & 10 & 4 & 32 & 20 & 25 & 19 \\
        C16C & m5 & 10 & 4 & 32 & 5 & 25 & 19 \\
        C17 & m5 & 10 & 8 & 32 & 10 & 25 & 19 \\
        C17B & m5 & 10 & 8 & 32 & 20 & 25 & 19 \\
        C17C & m5 & 10 & 8 & 32 & 5 & 25 & 19 \\
        C18 & m5 & 20 & 2 & 16 & 10 & 25 & 19 \\
        C18B & m5 & 20 & 2 & 16 & 20 & 25 & 19 \\
        C19 & m5 & 10 & 2 & 32 & 10 & 25 & 19 \\
        C19B & m5 & 10 & 2 & 32 & 20 & 25 & 19 \\
        C19C & m5 & 10 & 2 & 32 & 5 & 25 & 19 \\
        C20 & m5 & 12 & 2 & 8 & 10 & 25 & 19 \\
        C20B & m5 & 12 & 2 & 8 & 20 & 25 & 19 \\
        C21 & m5 & 12 & 2 & 9.5 & 10 & 10 & 4.75 \\
        C21B & m5 & 12 & 2 & 9.5 & 20 & 10 & 4.75 \\
        C21C & m5 & 12 & 3 & 12 & 10 & 10 & 4.75 \\
        C21D & m5 & 12 & 3 & 12 & 20 & 10 & 4.75 \\
        C22 & m5 & 40 & 2 & 8 & 10 & 25 & 19 \\
        C22B & m5 & 40 & 2 & 8 & 20 & 25 & 19 \\
        C22C & m5 & 40 & 2 & 8 & 5 & 25 & 19 \\
        C23 & m5 & 20 & 2 & 8 & 10 & 25 & 19 \\
        C23B & m5 & 20 & 2 & 8 & 20 & 25 & 19 \\
        C24 & m5 & 20 & 4 & 8 & 10 & 25 & 19 \\
        C24B & m5 & 20 & 4 & 8 & 20 & 25 & 19 \\
        C25 & m5 & 20 & 2 & 4 & 10 & 25 & 19 \\
        C25B & m5 & 20 & 2 & 4 & 20 & 25 & 19 \\

		\hline
	\end{tabular}

\section{Table of Characteristics of AWS EC2 Types}
\label{tab:characteristics_table}

	\centering
	\begin{tabular}{ccccccccccc} 
		\hline
		 Image & Per Hour & Per Hour  & Reserved & CPU     & Memory & Cores & Network   & EBS  
                   & Storage  \\
	   Type  &  Cost    & Spot Cost & Cost     & (Intel) &   (GB)     &       & Bandwidth & 
         Bandwidth  \\
		      &          &           &          &         &    &       & (Gbps)    & (Gbps)  &      \\
		\hline

        c6id.8xl & 1.61 & 0.7571 & 1.016 & Xeon 8375C (Ice Lake) & 64 & 32 & 12 & 10 & 1x1900 NVMe \\
        c5.24xl & 4.08 & 1.836 & 2.57 & Xeon Platinum 8275L & 192 & 96 & 25 & 19 & EBS \\
        m5.24xl & 4.608 & 2.163 & 2.903 & Xeon Platinum 8175 & 384 & 96 & 25 & 19 & EBS	\\
        m5.2xl & 0.384 & 0.1871 & 0.242 & Xeon Platinum 8175 & 32 & 8 & 10 & 4.75 & EBS \\
        c5id.24xl & 4.838 & 1.664 & 3.049 & Xeon 8375C (Ice Lake) & 192 & 96 & 37 & 30 & 4x1425 NVMe \\
		\hline
	\end{tabular}


\label{lastpage}
\end{document}